\documentstyle[version2,aps,preprint]{revtex}
\begin{document}

\begin{title}
An Asymptotic Solution of the $\infty-d$ Hubbard Model
\end{title}

\draft
\author{ X. Y. Zhang and  G. M. Zhang \cite{zhang2}}
\begin{instit}
International Center for Theoretical Physics, P. O. Box 586, 34100,
Trieste, Italy.
\end{instit}
\receipt{}

\maketitle
\begin{abstract}
We present an asymptotically exact solution of the $\infty-d$
Hubbard model at a special interaction strength $U_T$
corresponding to the strong-coupling Fermi-liquid fixed
point. This solution is intimately related to the Toulouse limit of
the single-impurity Kondo model and the symmetric Anderson model
in its strong-coupling limit.
\end{abstract}

PACS numbers: 71.30.+h, 75.10.Jm, 71.10.+x, 71.28.+d

\newpage
{\it Introduction:} Ever since the original introduction of the
$\infty-d$ Hubbard model by Metzner and Vollhardt \cite{mv}, and the
recognition that the large-$d$ limit model has a $k$-independent self-energy,
the hope for an exact solution has been high. After a series of important
analytical work \cite{mh}--\cite{vg}, and the recent surge in numerical
simulations of the
model, some of the outstanding physical processes described by the model have
been gradually made clear.
\cite{jrg}-\cite {gkkc}.
Yet, any exact solution is still out of sight.
The bottleneck is:
although the $k$-independent self-energy $ \Sigma(\omega) $ has
reduced the problem to a $0+1$ dimensional one with a self-consistency,
it is still highly non-trivial to calculate the Green's function of the
corresponding "impurity" problem, which is crucial for the self-consistency
to be complete.

Here, we present a solution which is exact in the asymptotic limit, {\it i.e.}
$ \tau\rightarrow\infty $. The solution is obtained only at half-filling
with a special interaction $U$.

We first sketch the lines of reasoning reaching this solution:
(i) The $\infty-d$ Hubbard model is mapped to a single impurity Anderson
model with a self-consistent condition \cite{gk}. The condition
requires
loosely speaking, that
the local Green's function to have the same form as
its surrounding conduction electron bath characterized by a
hybridization function.
(ii) As the interaction $U$ is increased, the dynamics of the impurity model
is governed by the "Kondo" physics, where there is an asymptotic
point, called the Toulouse limit (TL) \cite{tou}, at which the problem is
exactly
soluble. It will be shown that the corresponding symmetric Anderson Model also
 possesses
such a limit, where the asymptotic behavior of the local
Green's function can be obtained.
(iii) With this Green's function, the self-consistent procedure can
be carried out. It is found that, at the TL,
the asymptotic behavior of the impurity
Green's function can be made consistent with that of the hybridization
function, thus the $\infty-d$ Hubbard model is solved asymptotically.

{\it Toulouse limit of the symmetric impurity Anderson model:} It is well
known that the single-impurity Kondo model has such an asymptotic
limit: TL,
where the conduction electrons become free \cite{tou},
\cite{ayh}.
Similar limit exists for the Anderson model.
We first remind the readers how this limit is derived for the
Kondo model emphasizing on the concept rather than the details.

First, the Kondo Hamiltonian is separated into two parts,
$H= H_0 + H_I$, where
$ H_0 = {\sum}_{k,\sigma} \epsilon_{k} C_{k,\sigma}^{\dag} C_{k,\sigma} +
\frac{J_z}{2} S_z \sum_{k,k'}(C_{k,\uparrow}^{\dag}C_{k',\uparrow}-
C_{k,\downarrow}^{\dag} C_{k',\downarrow})$
and $H_I=J_{\perp}\sum_{k,k'}
(S^{+}C_{k,\downarrow}^{\dag}C_{k,\uparrow} +
S^{-}C_{k,\uparrow}^{\dag}C_{k,\downarrow}) $.
Because of the spin-flip
term $H_I$, a given spin sector of the system, say, the up spin,
constantly undergoes transition between the eigen states of
$H_i=\sum_k\epsilon_k C_{k,\uparrow}^{\dag} C_{k,\uparrow} +
\frac{J_z}{4} {\sum}_{k,k'}C_{k,\uparrow}^{\dag} C_{k',\uparrow} $, and
that of
$H_f=\sum_k\epsilon_k C_{k,\uparrow}^{\dag} C_{k,\uparrow} -
\frac{J_z}{4} {\sum}_{k,k'}C_{k,\uparrow}^{\dag} C_{k',\uparrow} $.
The two eigen states have a relative phase shift, $\delta =
2tan^{-1}{\frac{\pi\rho J_z}{4}}$, which is reflected in the propagator
between the two states,
$\langle|e^{iH_it}C_{\uparrow}e^{-iH_ft}C_{\uparrow}^{\dag}|\rangle\approx
t^{-(1-\delta/ \pi)^2}$.
This is the origin of the non-Fermi liquid behavior of the
X-ray edge problem \cite{nd}, \cite{ss}. When the phase shift
takes a particular value, such that the exponent in the asymptotic
Green's function becomes $ 2 ( 1 - \frac{\delta}{\pi} )^2 = 1 $, or
$ \rho J_z = \frac{4}{\pi}ctg(\frac{\pi}{2\sqrt{2}})$,
the Fermi-liquid (FL) behavior $\frac{1}{\tau}$ is restored \cite{note1},
and the system appears to be
free.  This is the origin of the TL. The restoring of the
FL behavior for an arbitrary phase shift $\delta$ is a much
more  subtle result first pointed out by Anderson {\it et al.} \cite{ayh}.

The symmetric  Anderson model
\begin{equation}
 H = \sum_{k,\sigma} \epsilon_{k} C_{k,\sigma}^{\dag}C_{k,\sigma} +
      V\sum_{k,\sigma}(C_{k,\sigma}^{\dag}f_{\sigma} + h.c.) +
      U n_{\uparrow}^{f} n_{\downarrow}^{f}
      -\frac{U}{2} ( n_{\uparrow}^{f} + n_{\downarrow}^{f} ),
\end{equation}
does not have a spin-flip term explicitly. But via a discrete
Hubbard-Stratonovich transformation \cite{hirsch}, we can write the partition
function as,
\begin{equation}
   Z = \sum_{\{\sigma_{l}=\pm 1\}} Tr\left( e^{-\Delta\tau
H(\sigma_1)}  e^{-\Delta\tau H(\sigma_2)}.....e^{-\Delta\tau
H(\sigma_L)}\right),
\end{equation}
where
$ \frac{1}{2}{\sum}_{\sigma} e^{\lambda \sigma( n_{\uparrow} - n_{\downarrow} )
 }
 = e^ { [-U n_{\uparrow}n_{\downarrow}
 +\frac{1}{2}(n_\uparrow+n_\downarrow) U ]\Delta\tau}$
and $ cosh\lambda=e^{(\frac{U}{2}\Delta\tau)} $.
Just like in the Kondo model, for a
given spin sector, say the up spin, the system is alternating between
eigen states of
$H_i=H_0+\frac{ \lambda}{\Delta\tau} f_{\uparrow}^{\dag} f_{\uparrow}$
($\sigma =1$)
and that of
$ H_f=H_0-\frac{\lambda}{\Delta\tau} f_{\uparrow}^{\dag} f_{\uparrow}$
($\sigma = -1 $). Here
$ H_0= {\sum}_k\epsilon_k C_{k,\uparrow}^{\dag} C_{k,\uparrow}  +
V{\sum}_k (C_{k,\uparrow}^{\dag} f_{\uparrow} + f_{\uparrow}^{\dag}
C_{k,\uparrow}) $ .
The phase shift of conduction electrons between these two eigen
states
is
$ \delta=2tan^{-1}(\frac{\pi V^2\rho_f}{{\frac{\lambda}{\Delta\tau}}})$
\cite{hamann}.

For a given sequence of $\{\sigma_l\} =++++----++++ $, where
the flips are taking place longer than the relaxation time
\cite{nd},\cite{ayh},\cite{hamann}, there
is a well-defined phase shift between the flipped states and the
unflipped ones.  This is reflected in the evolution
operator,
$u(\tau)= \langle|e^{H_i\tau}e^{-H_f\tau}|\rangle $.
The reason that it is an evolution operator instead of a Green's function
is because the spin-flip terms
are absent here.
This absence is compensated by the presence of the $f-$electrons
in the Hamiltonian, which will contribute to an overall phase shift
of $\pi$, so that
$\langle u(\tau)u^{\dag}(\tau')\rangle
\approx (\tau-\tau{'}) ^{-(1 - \frac{\delta}{\pi} )^2}$
still holds.

To arrive at the TL point,
the sum $\sigma$ is first
regrouped into sections that contain
equal number of spin-flips, $n$. Within each section, the positions of
the
spin-flips are summed, $\sum_{\sigma_1 \sigma_2 \sigma_3 ... \sigma_L }
= \sum_n \int\frac{ d\tau_1}{\Delta\tau} \int \frac{d\tau_2} {
\Delta\tau}...... $,
and the average is written as an exponential, \cite{ss}
 $$\langle u(\tau_1) u^{\dag} (\tau_2)u(\tau_3)....u^{\dag}(\tau_n)\rangle
 = e^{-(1 - \frac{\delta}{\pi})^2 \sum_{i>j}(-1)^{i+j} ln\mid
  \frac{\tau_i-\tau_j}{\Delta\tau}\mid }. $$
The partition function thus takes the form,
\begin{equation} \label {coulomb}
 Z = \sum_{n=0}^{\infty}
\int_{0}^{\beta}\frac{d{\tau}_{2n}}{\Delta\tau}
\int_{0}^{{\tau}_{2n}-\Delta\tau}\frac{d{\tau}_{2n-1}}{\Delta\tau}...
\int_{0}^{{\tau}_2-\Delta\tau}\frac{d{\tau}_1}{\Delta\tau}
 e^{-\left ( 2 ( 1 -\frac{ \delta}{\pi} )^2\sum_{i> j} (-1)^{i+j} ln\mid
\frac{({\tau}_i-{\tau}_j)}{\Delta\tau}\mid \right )}.
\end{equation}
Once the TL is set, {\it i.e.} the coefficient in front
of the logarithmic function becomes unity, the partition function is
just the same as that of the following Hamiltonian:
\begin{equation}\label {toulouse}
   H_{T} = \sum_{k} \epsilon_{k} C_{k}^{\dag}C_{k} +
      V_T\sum_{k}(C_{k}^{\dag}d + h.c.),
\end{equation}
where $V_T$ depends on the chemical potential of the Coulomb
gas which is ignored so far, but will be introduced in the following
calculations.
Since $H_T$ is free, the impurity Green's function $<d(t) d^{\dagger}(0)>$
can be easily obtained.

Notice that the same Coulomb gas formula (\ref {coulomb}) was derived
long time ago by Hamann \cite{hamann}
in his path integral approach.
His tunneling
configurations are the domain walls of the Ising variables
here.
That the symmetric Anderson model also has a TL
is not surprising, considering the fact that via Schrieffer-Wolff
transformation one can map it into a
Kondo model. The subtlety is, this transformation allows one to go to a
weak-coupling $J$ of the Kondo, whereas the TL is a
strong-coupling limit. The fact that the low-energy physics in the
weak-coupling limit is controlled by the strong-coupling fixed point,
keeps the physics of TL alive.

{\it Local Green's function of the symmetric impurity Anderson model:}
Comparing (1) to (4), the effect
of the $\sigma$'s sum is to push the
effective $f$ level to the Fermi point, {\it i.e.} zero, and to renormalize
the coupling parameters such that the low energy behavior of the
$f$-electrons is replaced by that of
spinless $d$-electrons. In fact, as pointed out by Anderson {\it
et al.} \cite{ayh},
$\langle n^{d}(t)n^{d}(0)\rangle =
\langle S_{z}(t)S_{z}(0)\rangle$.
This leads to a natural intuitive identification of $d$ with $f$ \cite
{tw}. One can express the low frequency part of the Green's function  as:
\begin{equation}\label{fdgf}
\langle f(t)f^{\dag}(0)\rangle_{low} \approx
 \frac{\Delta}{D} \langle d(t)d^{\dag}(0)\rangle,
\end{equation}
where $\Delta/D $ is the
spectral weight of the low frequency part. The weight of the high
frequency part
can be well approximated once the system is in the strong coupling
regime, and the bare band shape is known.  The sum of the two
is unity obeying the sum rule.

Another way of insuring the validity of
the above relationship, is to use the slave-boson
or slave-fermion decomposition
scheme \cite{rncaa}, to separate $f$-electrons as
$f_{\sigma}= a b_{\sigma} + \sigma d^{\dag} b_{-\sigma}^{\dag}.$
To constrain the $f$'s in a singly occupied state, we
restrict the bosons to $\sum_{\sigma}b_{\sigma}^{\dag} b _{\sigma}=1$.
It is then straight forwardly shown that
$S_z =\frac{1}{2} (b_{\uparrow}^{\dag} b_{\uparrow} - b_{\downarrow}^{\dag}
b_{\downarrow}) = b_{\uparrow}^{\dag}b_{\uparrow} -1/2, S^{\dag} =
b_{\uparrow}^{\dag} b_{\downarrow},
S^{-}=b_{\downarrow}^{\dag} b_{\uparrow} $, the usual Schwinger boson
representation of the spin operators. If we neglect the doublon part, then
we have
$<f_{\uparrow}(t)f_{\uparrow}^{\dag}(0)> = \langle  a(t)a^{\dag}(0)
b_{\uparrow}(t) b_{\uparrow}^{\dag}(0)\rangle\approx \langle a(t) a^{\dag}(0)
\rangle \langle  b_{\uparrow}(t) b_{\uparrow}^{\dag}(0)\rangle \propto
 \langle  b_{\uparrow}(t) b_{\uparrow}^{\dag}(0)\rangle $,
where one particle per site constraint has been enforced.
On the other hand, we
can write $ < d(t) d^{\dag}(0) > = < S^{-}(t) S^{+}(0) > $
using the spinless fermion representation of
the local spin operator.  We then replace the spin operator by
Schwinger bosons, such that,
$ \langle d(t) d^{\dag}(0) \rangle =
<b_{\downarrow}^{\dag}(t)b_{\uparrow}(t)b_{\uparrow}^{\dag}(0)
b_{\downarrow}(0)> \approx <b_{\uparrow}(t)b_{\uparrow}^{\dag}(0)>$. Thus,
we obtain eq.(\ref{fdgf}).

{\it An exact asymptotic solution of the $\infty-d$ Hubbard model:}
As pointed out by Georges and Kotliar \cite{gk}, the most fruitful way
to make use of a site-independent self-energy is to map the
model to an Anderson impurity model plus a self-consistency.
The impurity Lagrangian,
\begin{equation} \label {lagrangian}
   \pounds=-\int\int d\tau d{\tau}'C_{\sigma}^{\dag}G_{0}^{-1}C_{\sigma}+
       \int d\tau U(n_{\uparrow}-\frac{1}{2})(n_{\downarrow}-\frac{1}{2}),
\end{equation}
where
$G_{0}$ is the Green's function  with site $0$,
"the impurity site" removed. One first calculates $G$ from
$\pounds$, and relates $G$ to $G_{0}$ by,
$ G^{-1}(\omega)=G_{0}^{-1}(\omega)-\Sigma(\omega)$,
thereby closing up the self-consistency.
On a Bethe lattice, this relation is simplified to
$ G_{0}^{-1}=z-t^{2}G $,
The asymptotic form of $G_0$ can be obtained from its
spectral density representation,
$G_{0}(\tau)= {\int}_{0}^{\infty}
{\rho}_{0}(\epsilon)e^{-\epsilon\tau}d\epsilon $. At zero temperature,
$G_{0}(\tau)= \frac{{\rho}_{0}(0)}{\tau}+O(\frac{1}{{\tau}^{2}})$.
For $\infty -d$ Hubbard model, ${\rho}_{0}(0)$ is pinned \cite{mh},
because $\Sigma (0) = 0$, so that $  G ( 0 ) = G_0 ( 0 ) $. On a
Bethe lattice, the value of the pinning is $\frac{2}{\pi D}$,
where $D$ is the radius of the semi-circle density of states.

If we represent $G_0$ in terms of "integrated" conduction electrons
in the Anderson Model, as we are allowed to do in the case
of Bethe lattice,
$ G_{0}^{-1}=z- V^2 \Sigma \frac {1}{z - \epsilon_k }$, we can
transform (\ref {lagrangian}) to (1) with a conduction
electron density of states to be self-consistently determined.  In this way,
simply borrowing the last section can give the TL for the large-d Hubbard
model. But here,
we would like to derive this asymptotic limit in the Lagrangian formulation.
The final results are the same.

The partition function of
the effective impurity version of the $\infty -d$ Hubbard model can be
written as
\begin{equation}
Z= \sum_{\{\sigma_{l}=\pm 1\}} det ( G_0^{-1} +\frac{ \lambda}{\Delta\tau}
\sigma ) det ( G_0^{-1} -\frac{ \lambda}{\Delta\tau} \sigma ),
\end{equation}
where the same discrete Hubbard-Stratonovich transformation is used.
To obtain the Coulomb gas form, we make the following
expansions of the determinants:
\begin {equation} \label {det}
det \hat {O}_\sigma
det \hat {O}_{- \sigma}
= e ^{ - Tr ( G_0 \frac{\lambda}{\Delta\tau} \sigma ) ^2 + Tr O(G_0^4)...},
\end {equation}
where $\hat {O}_\sigma  =
I + G_0 \frac{\lambda}{\Delta\tau} \sigma $.
Since $G_0 \to  \frac{2}{\pi D}\frac{1}{\tau}$,
it is enough to keep the second order part only to obtain
the asymptotic limit.
Taking the trace
in the imaginary time,
$ Tr ( G_0 \sigma )^2 = \int d\tau_1 \int d\tau_2 G_0(\tau_1 -\tau_2 )
\sigma (\tau_2 ) G_0 (\tau_2 - \tau_1 ) \sigma (\tau_1 ) $
and using the fact $G_0 (\tau) = - G_0 (-\tau) $, we have
\begin {equation} \label {ising}
Z = e ^{  2 (\frac{2\lambda}{\pi D\Delta\tau} )^2 \sum_{i> j}
\sigma_i \frac {1}{ (i - j ) ^2 } \sigma_j  + \mu \sigma_i \sigma_{i+1}},
\end {equation}
where $\mu$ is the chemical potential
governed by the short time behavior of $G_0$ to be determined
below.
This long range 1-d Ising model can be mapped to a Coulomb gas model
\cite{ayh},
\cite{note2}.
The interaction strength in front of the logarithmics is
$2 (\frac{2\lambda}{\pi D\Delta\tau} )^2$.   When this is set to unity,
we obtain the TL of the $\infty -d $ Hubbard model:
$\sqrt{\frac{U}{\Delta\tau}}=\frac{\pi D}{2\sqrt{2}} $,
where we have approximated $\lambda \approx \sqrt{U\Delta\tau}$.
The low frequency part of the Green's function can be obtained
from eq.(\ref{toulouse}):
\begin{equation} \label {green}
  G(\omega)_{low}
    =\frac{\Delta / D }{ \omega - {V_T}^{2} \Sigma_{k}
    \frac{1}{\omega-{\epsilon}_{k}}}
    \approx \frac{\Delta / D }{\omega+i\Gamma}
\end{equation}
where $V_T = \frac{e^{- \mu } } {\Delta\tau} $,
and $\Gamma \equiv \pi V_{T}^2 \rho (0) = \frac{2}{D} V_{T}^2$.
Since $G ( 0 ) = \frac {2 } {i D} $, we have
$ \Delta = \frac { 4} {D \Delta\tau^2} e ^{-2\mu}, $ from which
$\Delta$ can be determined,
because $\mu$ is a function of $\Delta$.

Although the asymptotic solution $G (\tau ) = \frac{2}{\pi D} \frac{1}{\tau} $
does not depend on $\Delta$, neither does $U_T$, it is essential
that $\Delta $ being finite \cite {kotliar} for the TL to exist
as can be seen from eq.(\ref {green}).
To determine $\Delta $, we have to consider the short time behavior
or high energy part of $G_0$ which is lattice dependent and
cannot be calculated accurately within this framework.  Fortunately,
at least for the Bethe lattice, we can obtain an approximate value
of $\Delta$, which is finite at $U_T$, and thus proving the existence
of the TL.

At zero temperature, and for $\tau > 0$,
$G_0 (\tau ) = \int^{\infty}_{0} \rho_0 (\epsilon ) e^{-\epsilon\tau}
d\epsilon$
where $\rho_0 = - \frac{1}{\pi} Im G_0 $.
The TL point is a strong coupling point.  Our numerical experience
with the Bethe lattice in this region shows that $\rho_0$
consists of a $\delta-$like peak around $z_0 = \sqrt{\Delta D}$
in addition to the finite part $\frac{2}{\pi D}$ at $z=0$.
The former contribute to the short time behavior of $G_0 (\tau )$,
the latter to the long time behavior.
This has been derived analytically in Ref. \cite {rkz}.
For completeness, we rederive it briefly here.
On the Bethe lattice, $G_0 = \frac{1}{z -t^2 G} $.
The $G$ can be decomposed into a sum of
low and high energy parts,
$G = G_{low} + G_{high} $.
Here $G_{low}$ is the one obtained from eq.(\ref{toulouse}) and
$G_{high}$ can be written as
$ \int \frac{ \rho_h (\epsilon ) d\epsilon}
{\omega - \epsilon } $ where $\rho_h$ only counts for the high energy
part so that $\rho_h (\epsilon \to 0 ) \to 0 $.
For a small $\Delta$, in the region $\Delta < z < \sqrt{\Delta D} $,
$G_{high}(z) \approx
-z \int \frac{ \rho_h (\epsilon ) d\epsilon }{\epsilon^2} $,
and $G_{low}(z) \approx \frac{\Delta/D}{z} $.
Thus, we have
$G_0(z) \approx \frac{z}{z^2 (1 +C ) - t^2 \Delta/D }$
where $C = t^2 \int\frac{\rho_{h}d\epsilon}{\epsilon^2} $ and $t = D/2$.
The position of the $\delta$-like peak is found at,
$z_0 = \frac{1}{2} ( \frac{\Delta D}{1 + C} )^{\frac{1}{2}} $
with a weight of $\frac{1}{2(1+C)}$.
So,
for $\tau$ being finite, we have:
$G_0 (\tau ) \approx \frac{1}{2(1+C)} e^{-z_0\tau}$.
Using (\ref {det}) and (\ref {ising}), we obtain the
chemical potential from the short time part of $G_0$:
$\mu \approx \frac{\lambda^2}{4(1+C)^2} e^{-2 z_0 \Delta\tau } $.

So far, everything is dependent on the cut-off
$\Delta\tau$.  There are certain arbitrariness in choosing
this cut-off, which will affect the exact values of $U_T$ and
$\Delta$.  This is deeply rooted in the TL of the impurity
models.
Fortunately,
the main result of this paper, {\it i.e.} the form of the asymptotic
solution and the existence of this solution is independent of
the choice of the cut-off.
We adapt Hamann's choice of the cut-off
$\Delta \tau = \frac{6}{U} $ \cite{hamann}.
With this choice and an estimate of $C \approx t^2/(\frac{u}{2})^2 =
( \frac{D}{U_T} )^2 $, we
have, $U_T = 2.6 D $ and $\Delta = 0.2 D $.  The result is very
close to the second order perturbation calculation \cite {xyz},
although as we pointed out, the exact number comparison
is not very meaningful due to the cut-off dependence.

For a hypercubic lattice, the kinetic energy is more spread
out because of the Gaussian density of states in contrast to
the semi-circular one.  As a result, for the same value of interaction U,
$\Delta $ is expected to be larger.  Therefore, the TL
will also exist there although the calculation in determine $\Delta$
is much more complicated.

{\it Discussion:}
(i) We want to emphasize that the TL point obtained from the
Coulomb gas analogy is valid only asymptotically.  In the
Bethe Ansatz solution of the Anderson impurity model, there
is no special value of $U$ at which the rapidity equation
becomes free.  The same can be said about the Kondo model:
even though there is a TL-like limit in the Bethe
ansatz solution,
it is different from the one
derived by Toulouse \cite{tou} or Anderson {\it et al.} \cite{ayh}.
The difference lies in the phase shift.
The Toulouse line in the
Bethe ansatz solution intersects with the Wilson
fixed point \cite{tw}.
The phase shift of
the Wilson fixed point is $\pi$, the well-known unitary limit.
The phase shift of the TL point
is $(1-\frac{\sqrt{2}}{2})\pi$,
and is independent of the $J_{\perp}$.
In this sense, the usual TL
does not show up in the Bethe ansatz
for the Kondo problem either.

(ii) The approximation involved here
is to assume certain path being
important, {\it i.e.} the spin-flips are
separated long enough so that the system has a chance to relax.
In this way, the concept of phase shift can still be used.
In the Kondo regime,
the interval
of the adjacent spin-flip is about the order of the inverse Kondo temperature,
which is long compare to the relaxation time.
Therefore, this approximation becomes exact in the asymptotic
sense.

(iii) Within the above mentioned limitations, we have derived
an exact asymptotic point of the $\infty -d$ Hubbard model, which
corresponds to the strong-coupling FL fixed point.
The form of the solution itself does not provide any
new information about the model.
This is due to
a special feature of the large-d limit: the density of states is pinned
on the metallic side. As long as the FL assumption
is correct, one can arrive at this solution \cite{gk}.
But, the FL assumption can and does break down
when the interaction strength is large enough \cite{jrg}.
The assumption has to be self-consistent to be valid.  As stated
in the beginning, this is the most difficult part of the
$\infty-d $ Hubbard model.  The contribution of this work is
to show how the self-consistency can be achieved exactly at one point.
To generalize to arbitrary interaction strength,
we have also tried and failed to use another
asymptotic approaches for this problem, namely,
the conformal field theory
\cite {la}.

(iv) A more fruitful way of using the TL is to
go away from half-filling, where much less has been understood.
All the derivations used here are applicable except for one:
the distinction between the high energy part and the low energy
part spectral weight becomes blurred, thus making it more difficult to
determine the value of $\Delta $.

(v) Through out the derivations, we have emphasized on the phase shift.
Once the phase shift takes a special value, the model appears to
be free.  In this sense, X-ray edge problem also has a
TL \cite{np},
where the Anderson catastrophe
seemingly goes away. But unlike the X-ray edge problem where the phase
shift is well-defined between given initial
and final states, for the
single-impurity models and the $\infty-d$ Hubbard model, the phase shift
is only meaningful in an intermediate step of a special treatment:
the path-integral approach.  In this special language, the system
is phase-shifting back and forth until it is
equilibrated.  As creatively conjectured by Anderson {\it et al.}
\cite{ayh}, no matter what the phase shift is, as the system equilibrates,
it will only take one value, the TL.  This one statement
turns out to involve all the machinery of many-body physics to
justify and to modify \cite{wilnoz}. Thus, the essence of solving the
model at the TL directly is to reach the strong coupling point
without encountering the formidable task.
The present work is just one more example in exploiting this limit
\cite{leek}.

{\it Acknowledegement}.  We are grateful to Yu Lu for his
strong support and encouragements.  We want to
thank A. Georges, W. Krauth, Yu Lu
for many useful discussions.
One of us (X.Y.Z.) is indebted to G. Kotliar and P. Nozi\`{e}res
for their many critical comments
which helped to sharpen the arguments in the paper considerably,
and to the hospitality of the Theory Group of the Institute Laue-Langvine
where part of the work was completed.


\begin{references}
\bibitem{note1}the factor of two is due to the spin degeneracy.
\bibitem{kotliar}
G. Kotliar made a critical comment
that for the TL to exist, high energy part of the Green's function must
play a role.
\end{references}
\end{document}